# Ligand unbinding mechanisms and kinetics for T4 lysozyme mutants from τRAMD simulations

*Ariane Nunes-Alves[a,b], Daria B. Kokh[a], Rebecca C. Wade[a,b,c*]*

[a]Molecular and Cellular Modeling Group, Heidelberg Institute for Theoretical Studies, Schloss-Wolfsbrunnenweg 35, 69118 Heidelberg, Germany

[b]Center for Molecular Biology (ZMBH), DKFZ-ZMBH Alliance, Heidelberg University, Im Neuenheimer Feld 282, 69120 Heidelberg, Germany

[c]Interdisciplinary Center for Scientific Computing (IWR), Heidelberg University, Im Neuenheimer Feld 205, Heidelberg, Germany.

**Corresponding Author:** Rebecca.Wade@h-its.org

**ANA and DBK are joint first authors.**

**Abstract**:

The protein-ligand residence time, $\tau$, influences molecular function in biological networks and has been recognized as an important determinant of drug efficacy. To predict $\tau$, computational methods must overcome the problem that $\tau$ often exceeds the timescales accessible to conventional molecular dynamics (MD) simulation. Here, we apply the $\tau$-Random Acceleration Molecular Dynamics ($\tau$RAMD) method to a set of kinetically characterized complexes of T4 lysozyme mutants with engineered binding cavities. $\tau$RAMD yields relative ligand dissociation rates in good accordance with experiments and thereby allows a comprehensive characterization of the ligand egress routes and determinants of $\tau$. Although ligand dissociation by multiple egress routes is observed, we find that egress via the predominant route determines the value of $\tau$. We also find that the presence of metastable states along egress pathways slows down protein-ligand dissociation. These physical insights could be exploited in the rational optimization of the kinetic properties of drug candidates.

The residence time of a ligand-protein complex ($\tau$, given by the inverse of the dissociation rate: $1/k_{off}$) has become an important parameter in drug design, since for some targets, it shows a stronger correlation than the binding affinity with *in vivo* drug efficacy[1–4]. However, the determinants of protein-ligand residence times are not well understood. Moreover, the prediction of $\tau$ by molecular dynamics (MD) simulation is challenging, in particular because of the timescales involved. Although simulations of protein systems may now routinely extend to microseconds, they are short compared to typical values of $\tau$ for drug-like molecules. To overcome this problem, many computational methods to enhance sampling of ligand unbinding during MD simulations have been proposed[5,6]. However, it remains to be determined to what extent such approaches, which in some cases employ non-equilibrium perturbations, are able to correctly capture mechanistic details of ligand egress routes.

T4 lysozyme (T4L) mutants that contain engineered small artificial cavities that can accommodate benzene and indole derivatives have long served as model systems for investigating the fundamental mechanisms underlying protein-small molecule interactions and for benchmarking computational methods[7,8]. Remarkably, no less than thirteen computational studies have been published since 2018 by different research groups in which methods based on MD simulation were used to identify paths from a buried cavity to the T4L exterior and to try to characterize the ligand binding and unbinding processes energetically and kinetically (see **Figure 1**, review in Ref. [6] and references [9–21]).

Here, we apply the $\tau$RAMD[22,23] procedure to compute relative $\tau$ values for a set of T4L-ligand complexes with the goals of (1) assessing the ability of the procedure to compute accurate relative $\tau$ values, and (2) comprehensively characterizing the ligand egress routes and identifying the determinants of residence times. In the τRAMD procedure, relative τ values are computed from the ligand dissociation times observed in a set of random acceleration MD (RAMD) trajectories. In RAMD, a randomly oriented force is applied adaptively to the ligand during an MD simulation to enhance the rate of ligand unbinding. We find that, due to its computational efficiency and accuracy, τRAMD enabled a more complete characterization of T4L-ligand unbinding than has been reported in prior computational studies and that it yielded a good correlation between computed and experimental residence times for different ligands, mutants and environmental conditions. Furthermore, the mechanistic insights obtained from the τRAMD simulations allow us to understand how the presence of metastable states along ligand egress paths can affect residence times and how good estimates of protein-ligand residence times can be obtained without having to sample all the egress paths.

***τRAMD accurately predicts relative residence times for benzene and indole bound to T4L mutants.*** τRAMD was used to generate dissociation trajectories and to compute relative residence times (as described in Computational Methods in the Supporting Information) for indole and benzene bound to the buried cavity in the L99A mutant of T4L (T4L:L99A), and benzene bound to two additional T4L mutants: T4L:M102A and T4L:F104A (at 20 °C). Additionally, the dissociation of benzene from T4L:L99A was simulated at 10 and 30 °C to compare with the experimental measurements in Ref. [24].

The computed relative residence times show a remarkably good correlation with experimental data ($R^2 = 0.78$), with a mean unsigned error of about 38% of the experimental τ values, see **Figure 2**. Notably, τRAMD captures the trends in residence time despite the different determinants of τ, which does not correlate with the equilibrium dissociation constant for these systems (**Table S1**).

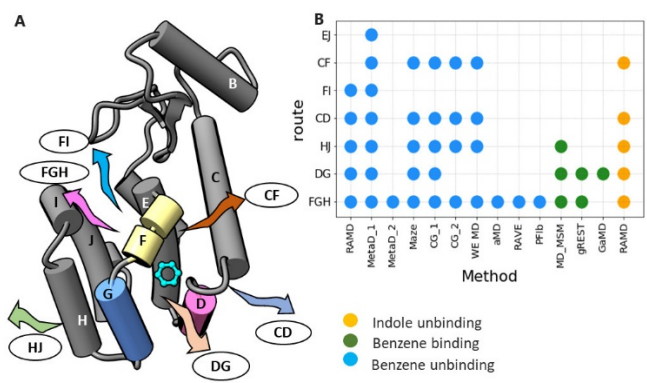

***Figure 1.*** *Egress routes observed for benzene and indole dissociation from the T4L:L99A mutant. (A) Cartoon representation of the protein with helices labelled. Egress routes are denoted by the helices lining them. (B) Ligand binding and unbinding routes observed in recent computational studies are indicated by circles colored by simulation type. Methods: MetaD_1[17] and MetaD_2[25] - metadynamics, Maze[18], CG_1[14] and CG_2[16] - coarse-grained, WE MD[9] – weighted ensemble MD, aMD[20] - accelerated MD, RAVE[13] - Reweighted autoencoded variational Bayes for enhanced sampling; PFIb[12] - Past-future information bottleneck, MD_MSM[21] – conventional MD and Markov State Model, gREST[11] - generalized replica exchange with solute tempering, GaMD[26] – Gaussian accelerated MD.*

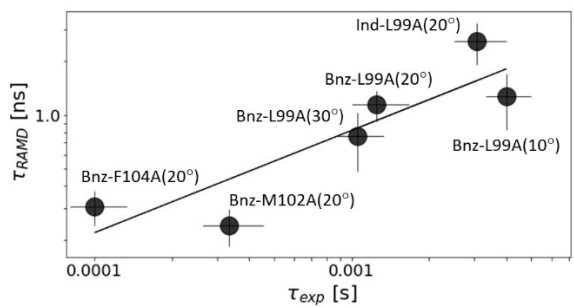

**Figure 2**. *Comparison of computed ($\tau_{RAMD}$) and measured ($\tau_{exp}$) residence times for benzene and indole for three T4L mutants at 10, 20 and 30 °C. Values (which are also given in **Table S1**) are plotted on the logarithmic scale and a linear fitting of computed to experimental data with $R^2$=0.78 is shown by the line. For benzene bound to T4L:F104A, $\tau_{exp} < 10^{-4}$ s and the error bar is defined as 25% (estimated from the uncertainty of the other measurements).*

**The populations of the unbinding paths depend on the binding site, ligand type and temperature.**

To explore the ligand egress pathways, we generated interaction fingerprints (IFPs) for the last 300 snapshots (i.e. 0.3 ns) of each RAMD trajectory, which generally encompass the last part of the ligand motion in its bound state and the complete ligand unbinding process (see Computational Methods in Supporting Information for details). We first extracted final frames with non-zero protein-ligand IFPs from the dissociation trajectories and carried out hierarchical clustering of these frames for each complex type.

For benzene dissociating from T4L:L99A, five egress routes were identified (shown in **Figure 3A** and **Figure S1**). These were also reported in most of the previous simulation studies (see **Figure 1**), with the FGH route (routes are named by the helices lining them) being clearly predominant at all conditions. The FGH route was the only one recorded in several publications using metadynamics, machine learning and aMD approaches[12,13,20,25]. However, the FGH route can be subdivided further by lowering the threshold for hierarchical clustering (for example, revealing the HJ route, see **Figure S4** and Ref. [17]). The two additional pathways, CF and EJ, were observed with low populations in several enhanced sampling simulations[9,14,17,18] but were not observed here for benzene although the CF pathway was observed for indole (see **Figure S5**).

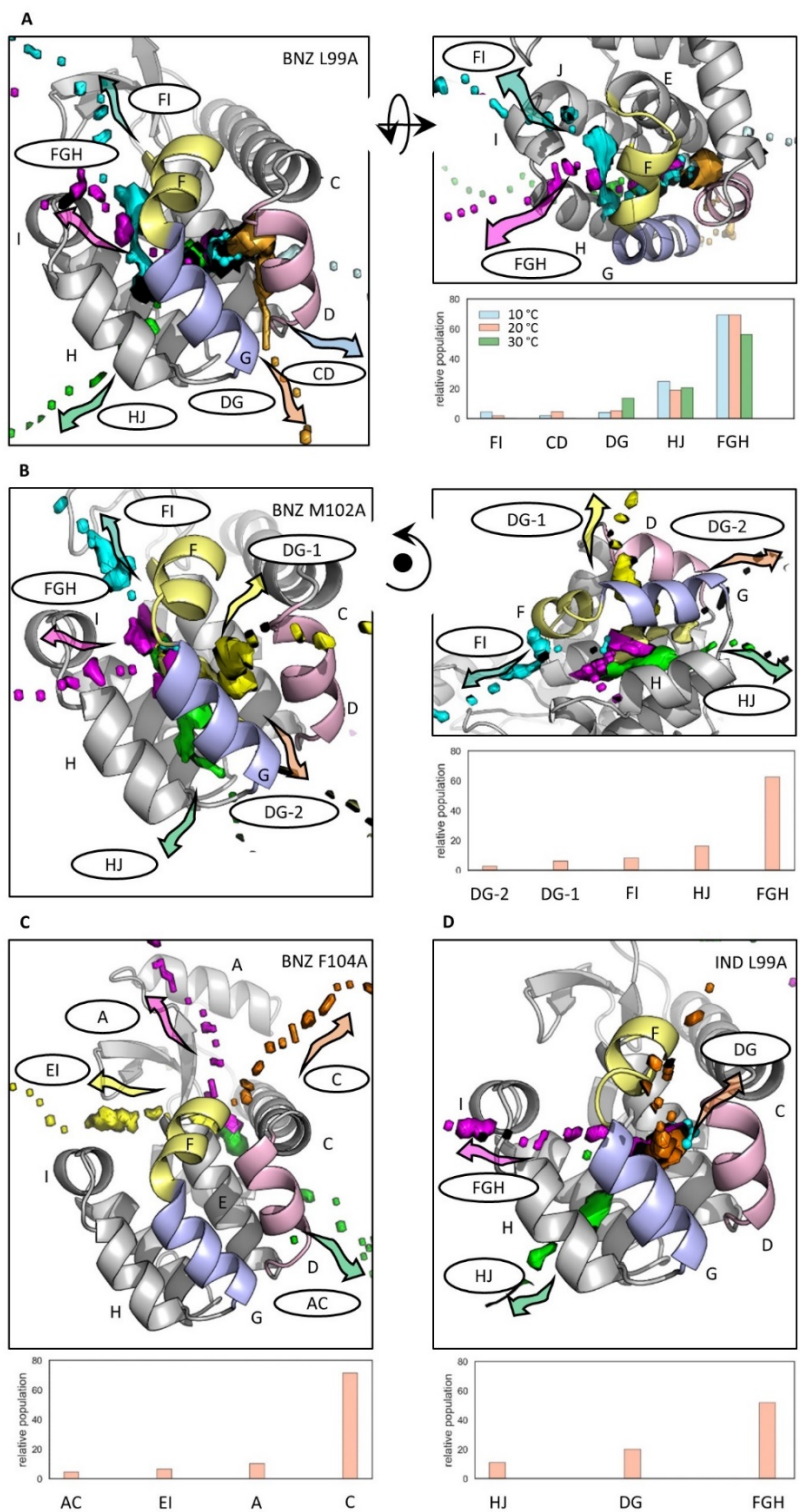

**Figure 3**. *Paths and their relative populations observed in RAMD trajectories for benzene dissociation from T4L:L99A at 10, 20 and 30 °C (A), benzene dissociation from T4L:M102A (B) and T4L:F104A (C),*

*and indole dissociation from T4L:L99A (D) at 20°C. The main dissociation paths observed were obtained from hierarchical clustering and are labeled according to the helices that they pass between. Each path is represented by one or two arbitrarily chosen dissociation trajectories (that belong to the corresponding cluster) displayed as isosurfaces of the population density obtained by mapping positions of the ligand center of mass in all frames of the trajectory onto a 3D grid. The directions of dissociation are indicated by arrows. The protein is shown in cartoon representation with helices labelled by letters and helices D, F and G colored pink, yellow and blue, respectively. The ligands are shown in cyan ball-and-stick representation in their bound position.*

The bound position of benzene in T4L:M102A is slightly shifted towards A102 (away from helix D) relative to its position in T4L:L99A. Accordingly, the egress routes are very similar to those for T4L:L99A: FGH, DG, HJ and FI, except for CD, which is not observed. The DG route can be further divided into DG-1 and DG-2, depending on whether the dissociation direction is perpendicular or parallel to the D and G helices (**Figures 3B and S2**).

In T4L:F104A, benzene occupies another binding site that is large and highly solvent-exposed. The four main paths found lead directly from the large open cavity that is lined by helices A, C and E. Only a few trajectories were observed in the opposite direction, via pathway AC (**Figures 3C and S3**).

The dissociation routes of indole from the same bound position in T4L:L99A as benzene are quite similar (**Figures 3D and S4**). However, path CD, which was observed for benzene, had a low population for indole, likely due to its larger size (**Figure S5**). Path FI was not identified for indole, but could be considered as part of the wider path FGH (**Figure S4**).

***Different egress routes have similar ligand dissociation times.*** Remarkably, despite the large differences in population, there are only small differences in the average dissociation times computed for all the pathways (see **Figures S1-4**). These results agree with other studies, in which similar preliminary kinetic rates[9] or unbinding times[18] were computed for the different paths. This suggests that a reasonably accurate prediction of the unbinding rate for benzene can be made without exhaustive pathway sampling, as long as the main path, FGH, is sampled, as done, for example, in Ref. [21] .

***Visiting multiple intermediate metastable states makes dissociation slower.*** Metastable states on the dissociation pathways for each complex were identified using k-means clustering of the IFPs computed for the last 300 frames of each dissociation trajectory (see Supporting Information for details).  The pattern of the metastable states for the benzene-T4L:L99A system is very similar for all temperatures simulated (**Figures 4A** and **S6**): there are two metastable states with relatively low populations at average RMSD values of about 10 Å (clusters 6 and 7; cluster 8 corresponds to the unbound state), which are intermediates on the dissociation paths FGH and HJ (**Figure 3A**), and there is a highly populated (often visited) metastable state 5 at an RMSD of ~ 5 Å where benzene is located close to the side-chain of M102. All other states (average RMSD < 3 Å, clusters 1 - 4 in **Figure 4A**) can be assigned to variations of the bound state. The dissociation flow can be described as ligand transitions from metastable state 4 to 5, 6, and then complete dissociation (gray arrows in **Figure 4A**). Direct transitions from the bound states 1 and 3 to the dissociated state (via paths CD and DG, respectively) and from 5 to 7 to dissociation (path HJ) are also observed, albeit with a lower probability.

The present results are consistent with previous conventional MD simulations of ligand binding accompanied by a Markov State Model analysis, where two main intermediate macrostates were identified[21]: MS1 between helices G and H, and MS2 covering the region between helices D and G, which is represented by clusters 1-4 close to the bound state in our simulations (see **Figure 4A**). MS1, an intermediate on the main association pathway, is more spread out and includes regions occupied by clusters 5, 6 and 7. MS1 is the intermediate state with the most flux during ligand binding, while the alternative binding path through MS2 (analogous to dissociation route DG from cluster 1) was observed much less, in agreement with our analysis. Remarkably, although cluster 6 is an intermediate on the main dissociation flow (see path FGH in **Figure 3A** and the dissociation network in **Figure 4A**), it is slightly less populated and more spatially localized than cluster 7. Gating by F114 on the dissociation pathway is likely to be the main reason for the ligand spending time in cluster 6. However, flipping of F114 is rather fast and thus does not slow down ligand dissociation significantly, whereas squeezing between helices H and J (i.e. via metastable state 7) is notably slower, making dissociation path HJ less populated.

Enhanced sampling simulations of benzene binding by GaMD[26] and gREST[11] revealed only the metastable states between helices D and G (i.e. clusters 1-4). In unbinding simulations using aMD[20], intermediate states between helices G and H (cluster 5) and between helices F and H (resembling cluster 6 but shifted closer to helix H) were identified. Thus, not all of the metastable states identified in conventional MD and RAMD simulations were revealed by these enhanced sampling methods. Moreover, small changes in protein structure were associated with ligand unbinding (see **Figure S7**).

The dissociation flow for indole has a similar pattern to benzene in RAMD trajectories, albeit with a larger variety of bound states (clusters 1-5, **Figure 4D**). Indeed, the indole residence time in T4L:L99A at 20 °C is comparable to that for benzene at 10 °C. The temperature difference is consistent with indole being slightly larger and needing to squeeze through the narrow channel gated by F114.

In contrast to T4L:L99A, for the M102A and F104A mutants, the pattern of benzene dissociation trajectories is different: the main flow leads either directly from the bound state to dissociation (T4L:F104A, **Figure 4C**) or through the intermediate state located in the vicinity of the bound one (T4L:M102A, **Figure 4B**; cluster 5 is close to F114). Accordingly, the dissociation time of benzene from both mutants is notably shorter than from T4L:L99A.

Thus, the benzene residence times are related to the number of intermediate metastable states for egress from the three mutants in the RAMD simulations (**Figure 4A-C**). Each metastable state can be associated with a subsequent transition barrier along the dissociation path, and therefore, a corresponding prolongation of the dissociation time.

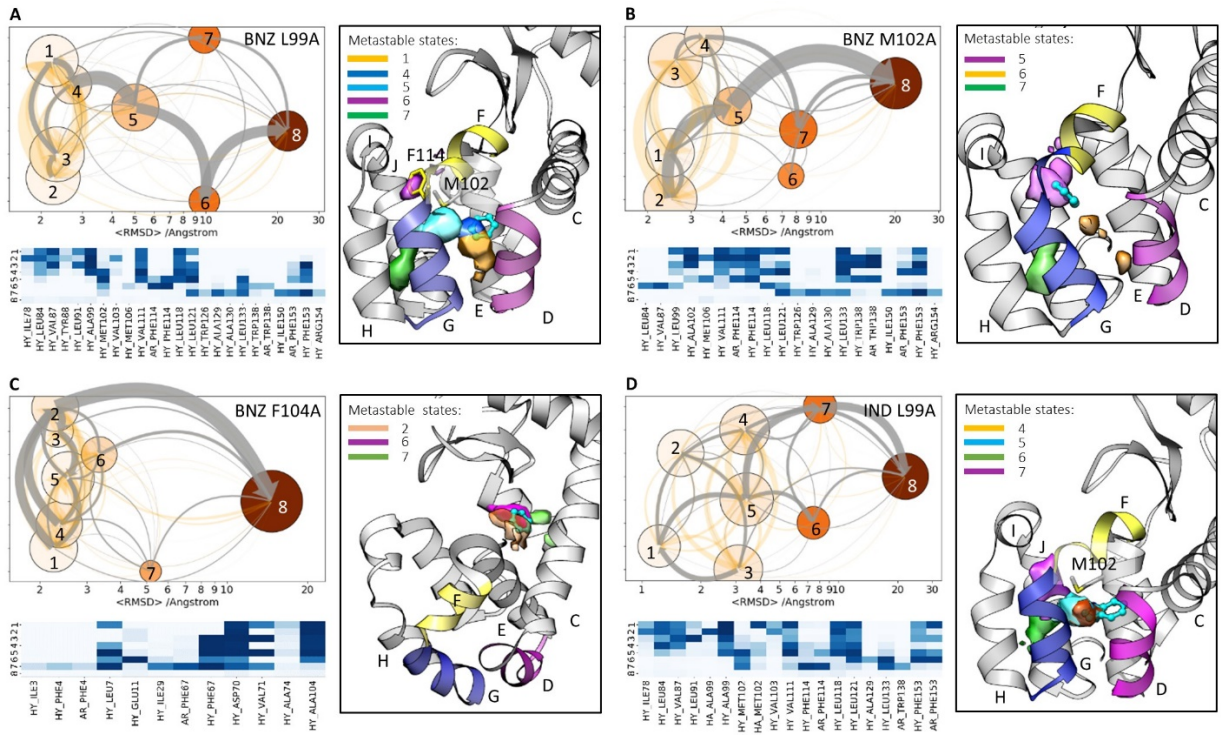

**Figure 4.** *Analysis of benzene unbinding from T4L:L99A (A), T4L:M102A (B) and T4L:F104A (C), and indole unbinding from T4L:L99A (D) in RAMD trajectories. Clusters were defined by clustering of frames from egress trajectories in IFP space. Dissociation pathways are shown in a graph-representation; each node represents a cluster or metastable state that is colored and placed according to increasing mean RMSD of the ligand in the cluster from in the starting complex; the node size denotes the cluster population; transitions between nodes are indicated by arrows for simulations at 20 °C (transitions for simulations at 10 and 30 °C are shown in* **Figure S6**)*: the net transition flux between nodes is shown by gray arrows with their thickness proportional to the flux magnitude; the transitions between states are shown by orange arrows with their thickness proportional to the transition number. Some clusters are displayed as isosurfaces of the ligand center of mass population density mapped onto the 3D grid. Helices D, F and G are shown in pink, yellow and blue, and the ligand is shown in cyan ball-and-stick representation. The heat maps show the composition of the clusters, in terms of ligand-protein contacts (color pallet from white to dark-blue indicates increasing contribution): HY: hydrophobic interactions. AR: aromatic interactions.*

In conclusion, we have presented a computational characterization of the unbinding processes for a set of complexes of T4L mutants with benzene and indole. We find that τRAMD provides very good agreement between computed relative unbinding rates and experimental data obtained for distinct conditions: different ligands, different mutants and different temperatures. We find that for benzene dissociation from T4L:L99A, there is one dominant egress path, FGH, which explains why only this pathway was found in many computational studies and indicates that accurate dissociation rates can be computed for this system even if just the main egress route is sampled. Our study also showed that longer τ is associated with more complex dissociation pathways with multiple intermediate metastable states (as seen for indole and benzene dissociating from T4L:L99A), in contrast to the one-step dissociation observed for complexes with shorter τ (benzene – T4L:M102A and T4L:F104A). The physical insights revealed here can be used in the rational optimization of the kinetic properties of drug candidates.

**Acknowledgements**

This work was supported by a Capes-Humboldt postdoctoral scholarship to A N-A (Capes process number 88881.162167/2017-01), the European Union's Horizon 2020 Framework Programme for Research and Innovation under Grant Agreements 785907 and 945539 (Human Brain Project SGA2 and SGA3), and the Klaus Tschira Foundation.

**References**

[1]    J. Romanowska, D. B. Kokh, J. C. Fuller, R. C. Wade, in *Thermodyn. Kinet. Drug Bind.* (Ed.: (eds G. M. Keserü and D. C. Swinney)), Wiley-VCH Verlag GmbH & Co. KGaA, Weinheim, Germany, Weinheim, **2015**, pp. 211–235.

[2]    M. Bernetti, M. Masetti, W. Rocchia, A. Cavalli, *Annu. Rev. Phys. Chem.* **2019**, *70*, 143–171.

[3]    R. A. Copeland, *Nat. Rev. Drug Discov.* **2016**, *15*, 87–95.

[4]    D. A. Schuetz, W. E. A. de Witte, Y. C. Wong, B. Knasmueller, L. Richter, D. B. Kokh, S. K. Sadiq, R. Bosma, I. Nederpelt, L. H. Heitman, E. Segala, M. Amaral, D. Guo, D. Andres, V. Georgi, L. A. Stoddart, S. Hill, R. M. Cooke, C. De Graaf, R. Leurs, M. Frech, R. C. Wade, E. C. M. de Lange, A. P. IJzerman, A. Müller-Fahrnow, G. F. Ecker, *Drug Discov. Today* **2017**, *22*, 896–911.

[5]    N. J. Bruce, G. K. Ganotra, D. B. Kokh, S. K. Sadiq, R. C. Wade, *Curr. Opin. Struct. Biol.* **2018**, *49*, 1–10.

[6]    A. Nunes-Alves, D. B. Kokh, R. C. Wade, *Curr. Opin. Struct. Biol.* **2020**, *64*, 126–133.

[7]    A. E. Eriksson, W. A. Baase, J. A. Wozniak, B. W. Matthews, *Nature* **1992**, *355*, 371–373.

[8]    A. E. Eriksson, W. A. Baase, X. J. Zhang, D. W. Heinz, M. Blaber, E. P. Baldwin, B. W. Matthews, *Science.* **1992**, *255*, 178–183.

[9]    A. Nunes-Alves, D. M. Zuckerman, G. M. Arantes, *Biophys. J.* **2018**, *114*, 1058–1066.

[10]   Y. Wang, O. Valsson, P. Tiwary, M. Parrinello, K. Lindorff-Larsen, *J. Chem. Phys.* **2018**,

*149*, 072309.

[11]    A. Niitsu, S. Re, H. Oshima, M. Kamiya, Y. Sugita, *J. Chem. Inf. Model.* **2019**, *59*, 3879–
        3888.

[12]    Y. Wang, J. M. L. Ribeiro, P. Tiwary, *Nat. Commun.* **2019**, *10*, 3573.

[13]    J. M. Lamim Ribeiro, P. Tiwary, *J. Chem. Theory Comput.* **2019**, *15*, 708–719.

[14]    B. R. Dandekar, J. Mondal, *J Phys Chem Lett* **2020**, *11*, 5302–5311.

[15]    S. D. Lotz, A. Dickson, **2020**, DOI 10.26434/chemrxiv.12801125.v1.

[16]    P. C. T. Souza, S. Thallmair, P. Conflitti, C. Ramírez-Palacios, R. Alessandri, S. Raniolo,
        V. Limongelli, S. J. Marrink, *Nat. Commun.* **2020**, *11*, 3714.

[17]    R. Capelli, P. Carloni, M. Parrinello, *J. Phys. Chem. Lett.* **2019**, *10*, 3495–3499.

[18]    J. Rydzewski, O. Valsson, *J. Chem. Phys.* **2019**, *150*, 221101.

[19]    J. Rydzewski, *Comput. Phys. Commun.* **2020**, *247*, 106865.

[20]    V. A. Feher, J. M. Schiffer, D. J. Mermelstein, N. Mih, L. C. T. Pierce, J. A. McCammon,
        R. E. Amaro, *Biophys. J.* **2019**, *116*, 205–214.

[21]    J. Mondal, N. Ahalawat, S. Pandit, L. E. Kay, P. Vallurupalli, *PLOS Comput. Biol.* **2018**,
        *14*, e1006180.

[22]    D. B. Kokh, M. Amaral, J. Bomke, U. Grädler, D. Musil, H.-P. Buchstaller, M. K. Dreyer,
        M. Frech, M. Lowinski, F. Vallee, M. Bianciotto, A. Rak, R. C. Wade, *J. Chem. Theory*

*Comput.* **2018**, *14*, 3859–3869.

[23] D. B. Kokh, T. Kaufmann, B. Kister, R. C. Wade, *Front. Mol. Biosci.* **2019**, *6*, 36.

[24] V. A. Feher, E. P. Baldwin, F. W. Dahlquist, *Nat. Struct. Biol.* **1996**, *3*, 516–521.

[25] Y. Wang, E. Papaleo, K. Lindorff-Larsen, *Elife* **2016**, *5*, e17505.

[26] Y. Miao, V. A. Feher, J. A. McCammon, *J. Chem. Theory Comput.* **2015**, *11*, 3584–3595.

**Entry for the Table of Contents**

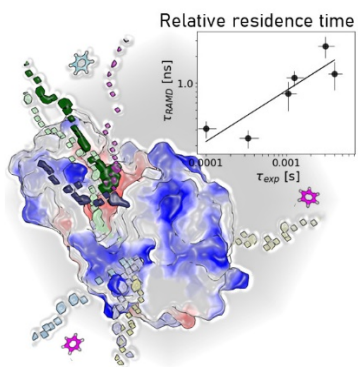

Efficient enhanced molecular dynamics simulation procedure, τ-Random Acceleration Molecular Dynamics (τRAMD), yields ligand residence times in good accordance with experiments for T4 lysozyme mutants, reveals ligand egress routes and mechanisms, and shows that visiting more metastable states slows protein-ligand dissociation.

**Keywords:** biophysics, molecular dynamics, proteins, ligand-protein residence time, ligand unbinding pathway

**Supporting Information**

# Ligand unbinding mechanisms and kinetics for T4 lysozyme mutants using τRAMD

*Ariane Nunes-Alves[a,b], Daria B. Kokh[a], Rebecca C. Wade[a,b,c*]*

[a]Molecular and Cellular Modeling Group, Heidelberg Institute for Theoretical Studies, Schloss-Wolfsbrunnenweg 35, 69118 Heidelberg, Germany

[b]Center for Molecular Biology (ZMBH), DKFZ-ZMBH Alliance, Heidelberg University, Im Neuenheimer Feld 282, 69120 Heidelberg, Germany

[c]Interdisciplinary Center for Scientific Computing (IWR), Heidelberg University, Im Neuenheimer Feld 205, Heidelberg, Germany.

**Corresponding Author**

*Rebecca.Wade@h-its.org

# Computational Methods

## A. Molecular dynamics simulations

The structures of the complexes of benzene bound to the T4L:L99A, T4L:M102A and T4L:F104A mutants were obtained from the PDB files 3HH4[27], 220L[28] and 227L[28], respectively. The crystal structure of indole bound to T4L:L99A was obtained from PDB file 185L[29]. Protonation states were assigned using pdb2pqr[30,31] at pH 5.5 to mimic kinetic experiments[24]. Crystallographic water molecules were maintained and crystallization molecules were removed.

The τRAMD method[22] was used to compute relative residence times. The protocol for the setup of each system and for molecular dynamics simulation was described in detail previously[22] and is therefore described briefly here. The AMBER ff14SB force field[32] was used for the protein and the GAFF force field[33] for the ligand. RESP[34,35] partial atomic charges for ligands were obtained using molecular electrostatic potentials from quantum mechanical calculations performed at the HF level with HF/6-31G** basis set using GAMESS[36]. Each system was solvated in a periodic box of TIP3P water molecules with a distance of 10 Å from the solute to the box edge using tleap. 7 $Na^+$ ions and 16 $Cl^-$ ions were added to achieve neutrality and an approximate ionic strength of 50 mM, which mimics the conditions of the kinetic experiments. The system was energy minimized and gradually heated to 293 K in 1 ns with the Langevin thermostat and harmonic restraints of 50 kcal/(mol $Å^2$) on all non-hydrogen atoms of the protein and the ligand using the AMBER14 software[37]. Pressure was adjusted to 1 atm in a further 2 ns simulation using the Berendsen barostat. Then the system was run for 2 more nanoseconds without restraints and the final snapshot was used as input for heating and equilibration simulations carried out with the NAMD software[38]. Heating from 0 to 293 K was performed for 6 ns in the NVT ensemble using the Langevin thermostat. Then equilibration was performed for 20 ns in the NPT ensemble using the Langevin thermostat and the Nosé–Hoover barostat for temperature (293 K) and pressure (1 atm) control, respectively. Eight replicas of equilibration simulations were performed for each system. The last snapshot of each replica was used to simulate ligand dissociation. For this purpose, the RAMD method[39] was applied, in which an additional force with a magnitude of 4 kcal/(mol Å) and a random direction was applied to the center of mass (COM) of the ligand in MD simulations. Every 100 fs, the force direction was changed randomly if the ligand COM did not move further than 0.025 Å and was retained otherwise. For each of the eight equilibration replicas, 15 RAMD dissociation trajectories were generated, resulting in a total of 120 trajectories for each system. Simulations were stopped and the ligand considered to be dissociated when the distance between the COMs of the ligand and the protein was greater than 40 Å. The time required for ligand dissociation in each trajectory was stored and the recorded dissociation times for the set of trajectories were used to compute the relative residence times.

**B. Analysis protocol**

The protein-ligand interaction fingerprints, IFPs, were computed for the last 300 frames saved at intervals of 1 ps of each RAMD trajectory using the MD-IFP method described in ref. [40]. The results were similar when the last 500 snapshots were used (**Figure S8**). The IFPs analysed for benzene and indole included aromatic interactions, hydrogen bonds, and hydrophobic interactions. Additionally, the coordinates of the ligand COM and the ligand RMSD from the starting equilibrated complex were stored for each frame. The computed IFPs for the last 300 frames of each trajectory were combined in one IFP binary table for each complex (the simulations of dissociation of the benzene-T4L:L99A complex at different temperatures were analysed together), with an entry of 1 if a contact was observed and zero if it was not observed. All frames were then split into 8 clusters based on their IFP composition by using a k-means algorithm, which provided the most populated states of the ligand in the IFP space (The simulations of the benzene-T4L:L99A complex at all three temperatures were analyzed together). Each cluster represents either the bound, a metastable, or the unbound state, which can be distinguished by the average RMSD of the ligand non-hydrogen atoms from their position in the starting complex. The position of the ligand COM mapped onto a 3D grid (with a spacing of 1 Å) was used to generate and display the COM density distribution for each cluster.

Additionally, the last frames of each RAMD dissociation trajectory that had IFP vectors containing at least 2 protein-ligand contacts were combined into a dissociation IFP set that was then clustered using a hierarchical clustering procedure. For each system, a variable clustering threshold was chosen in order to obtain 3-5 clusters corresponding to the main unbinding paths. The dissociation paths were then displayed by plotting the ligand COM density distribution for all trajectories or one representative trajectory in the corresponding cluster.

**Table S1.** *Experimental values of $K_D$, $k_{off}$ and $k_{on}$ for the T4L-ligand complexes studied from Ref.* [24]. *Comparison of these values shows that the mechanisms determining the length of the residence times differ among the systems studied.*[a]

| T4L mutant | Ligand | Temperature (°C) | $K_D$ (mM) | $k_{off}$ (s$^{-1}$) | $k_{on}$ (10$^6$ M$^{-1}$s$^{-1}$)[b] | $\tau_{RAMD}$ [ns] |
|---|---|---|---|---|---|---|
| L99A | | 10 | 0.3±0.2 | 250±50 | 0.83 | 1.27±0.44 |
| | | 20 | 0.8±0.12 | 800±200 | 1.00 | 1.15±0.22 |
| | Benzene | 30 | 1.1±0.11 | 950±200 | 0.86 | 0.75±0.27 |
| M102A | | 20 | 0.8±0.1 | 3000±800 | 3.75 | 0.24±0.05 |
| F104A | | 20 | 1.3±0.05 | >10$^4$ | >7.7 | 0.31±0.06 |
| L99A | indole | 20 | 0.35±0.06 | 325±75 | 0.92 | 2.60±0.68 |

[a]*The difference in $\tau$ between benzene and indole for binding to T4L:L99A at 20 °C can be ascribed solely to a relative stabilization of the bound state in the indole complex as both $K_D$ and $k_{off}$ differ by about the same factor (2.3). In contrast, the difference in $\tau$ for benzene binding to T4L:L99A and to T4L:M102A can be ascribed solely to lowering of the transition barrier for unbinding since the $K_D$ is the same for both mutants while the $k_{off}$ differs by a factor of 3.75. On the other hand, the difference in $\tau$ for benzene binding to the more exposed cavity of T4L:F104A versus the other mutants appears to be due to both a destabilization of the bound state and a lowering of the transition barrier.*

[b] *Computed as $k_{on} = k_{off}/K_D$*

# Supplementary Figures

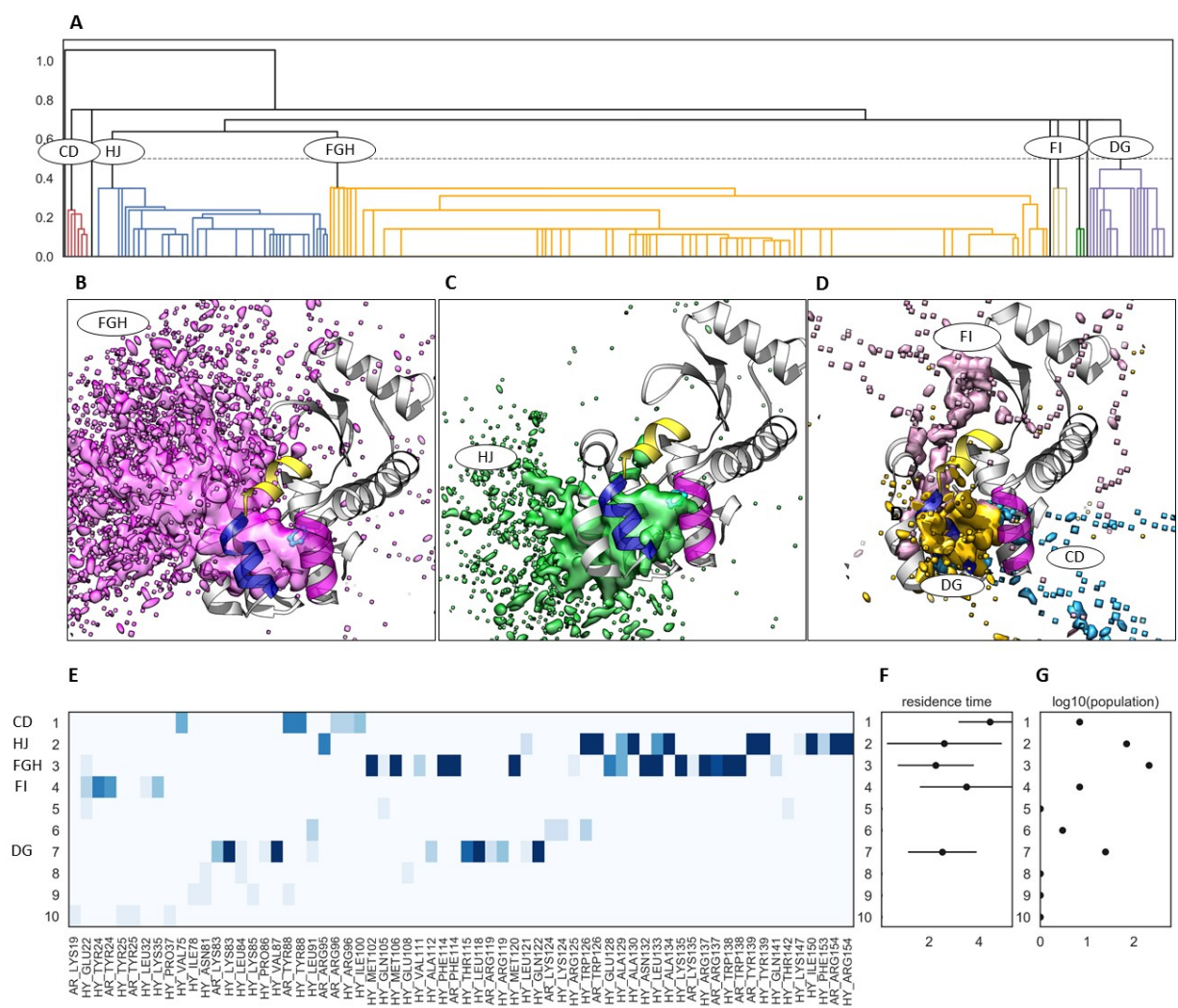

**Figure S1.** *Analysis of egress routes for benzene dissociation from T4L:L99A in RAMD simulations at three temperatures: 10, 20 and 30 °C. (A) Results of hierarchical clustering. (B-D) Display of the most populated clusters (populated by more than one snapshot (each representing one trajectory)). (E) Cluster composition in terms of IFPs. (F) Computed residence times (in nanoseconds) for the highly-populated clusters shown in (E). (G) The corresponding cluster populations shown on a logarithmic scale.*

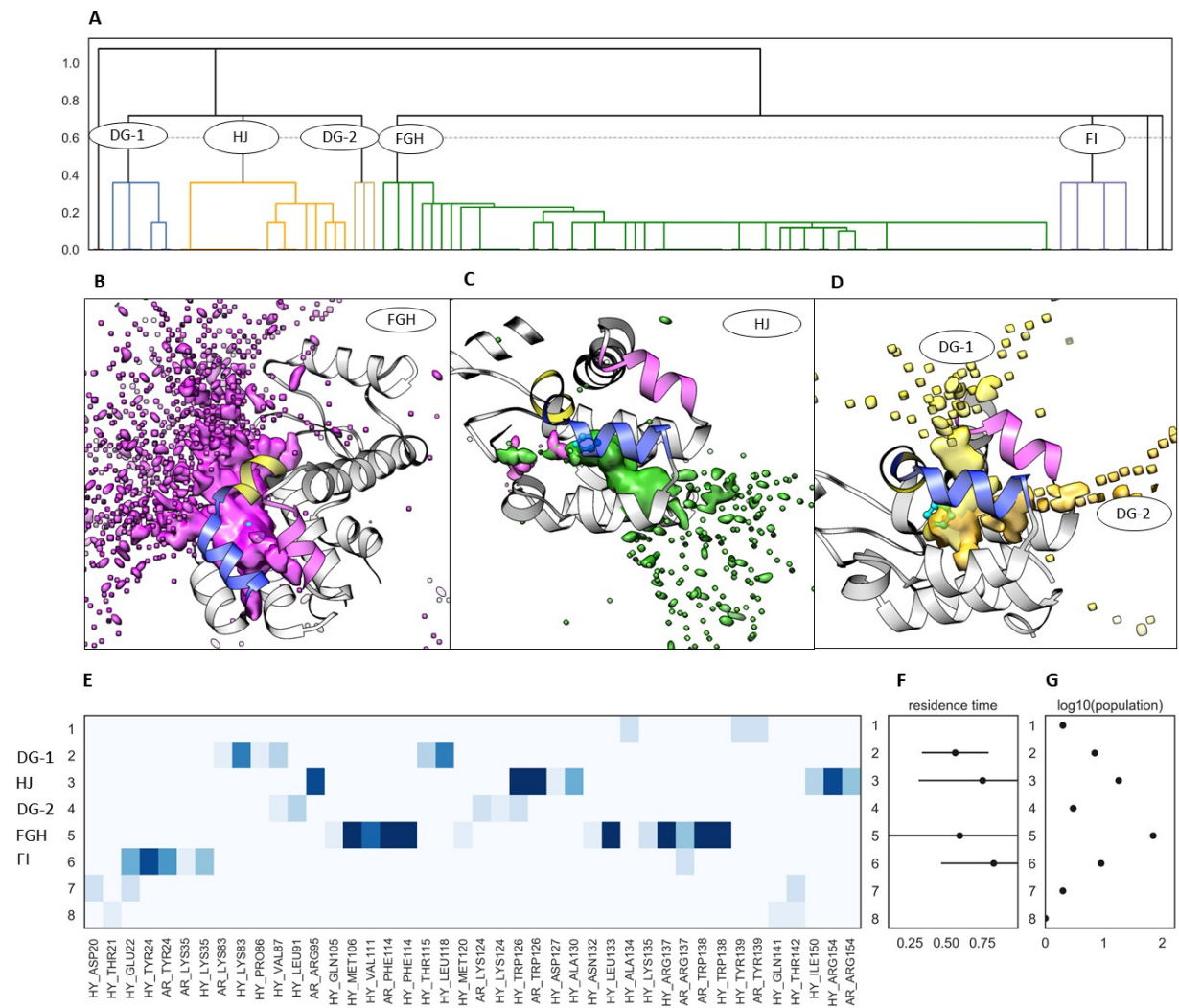

***Figure S2.*** *Analysis of egress routes for benzene dissociation from T4L:M102A. Legend as for Figure S1.*

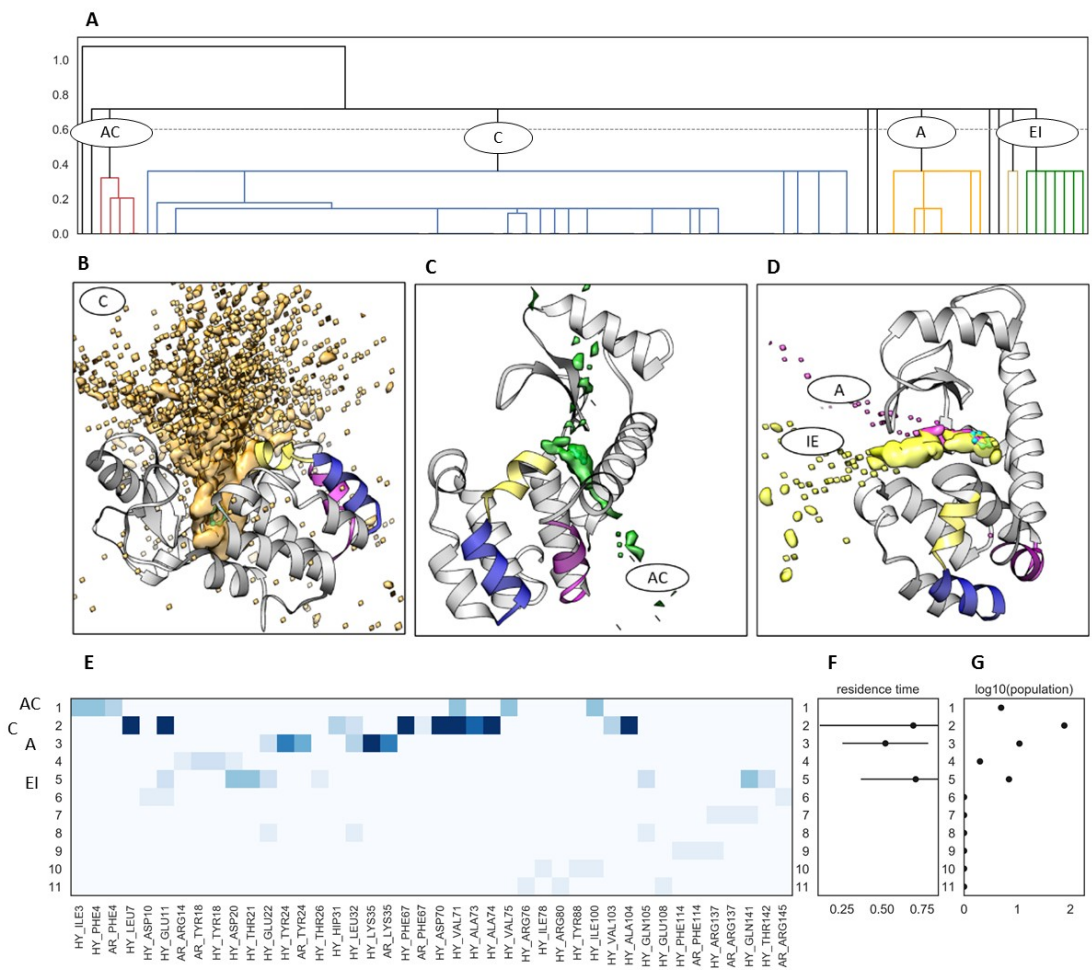

***Figure S3.*** *Analysis of egress routes for benzene dissociation from T4L:F104A. Legend as for Figure S1.*

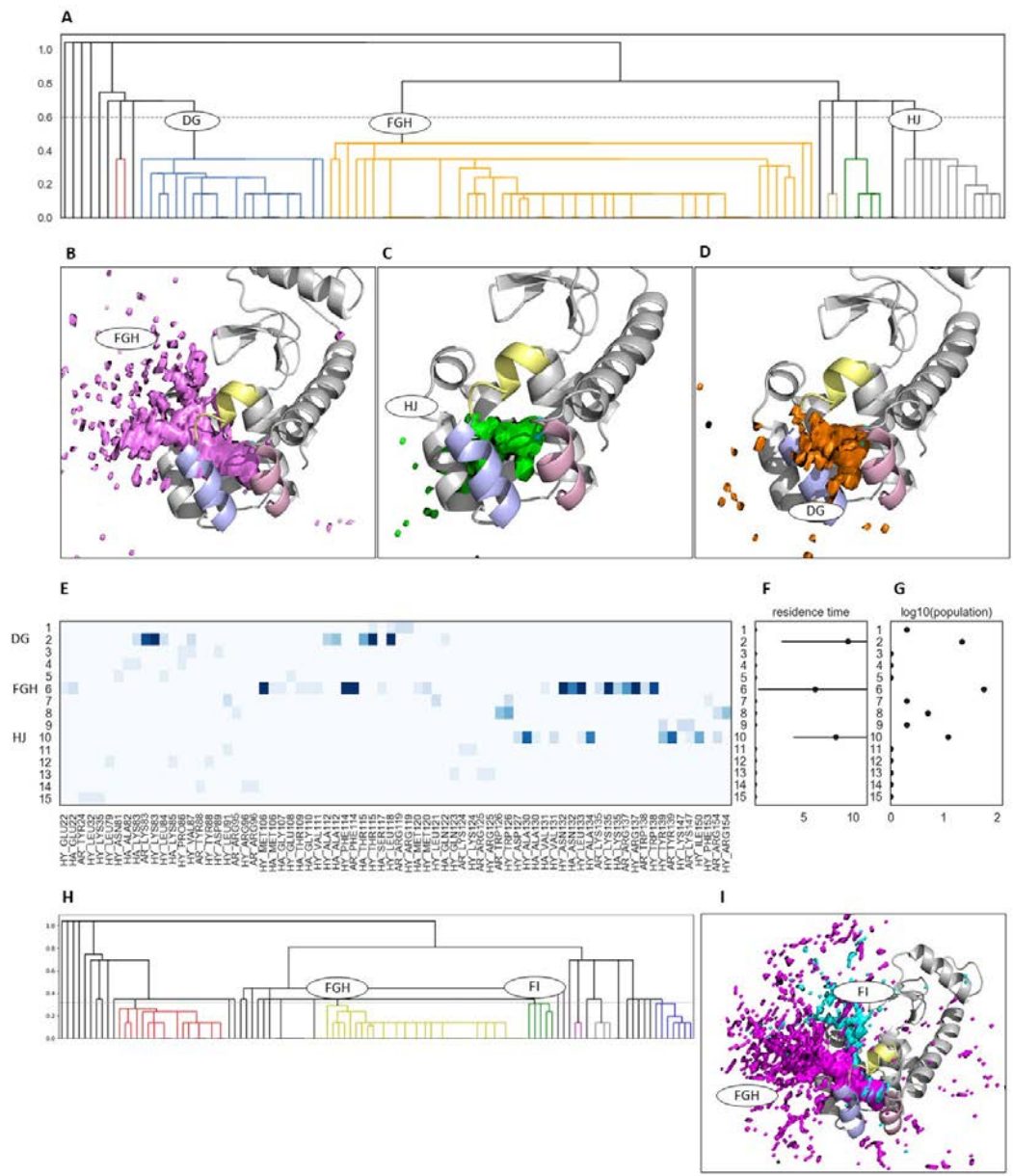

**Figure S4.** *Analysis of egress routes for indole dissociation from T4L:L99A. (A-G) Legend as for Figure S1. (H-I) Results for a smaller cutoff in the hierarchical clustering. (H) Results of hierarchical clustering. (I) Display of clusters for paths FGH and FI.*

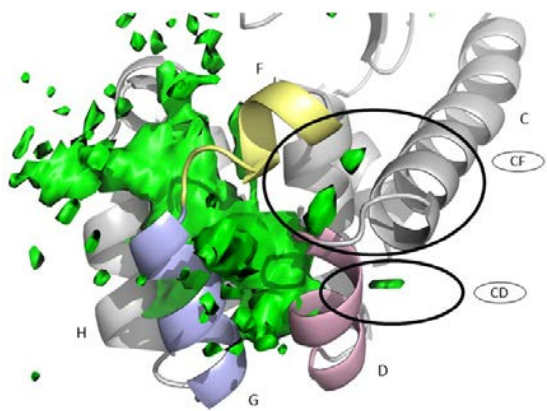

**Figure S5.** *All dissociation paths (green) for indole from T4L:L99A are displayed. Paths CD and CF have a low population (< 5%) and are therefore not shown in figure 3 of the main text. The paths are represented by an isosurface of the population density obtained by mapping the positions of the ligand center of mass (COM) in all frames of the trajectories onto a 3D grid. Helices D, F and G are shown in pink, yellow and blue, respectively.*

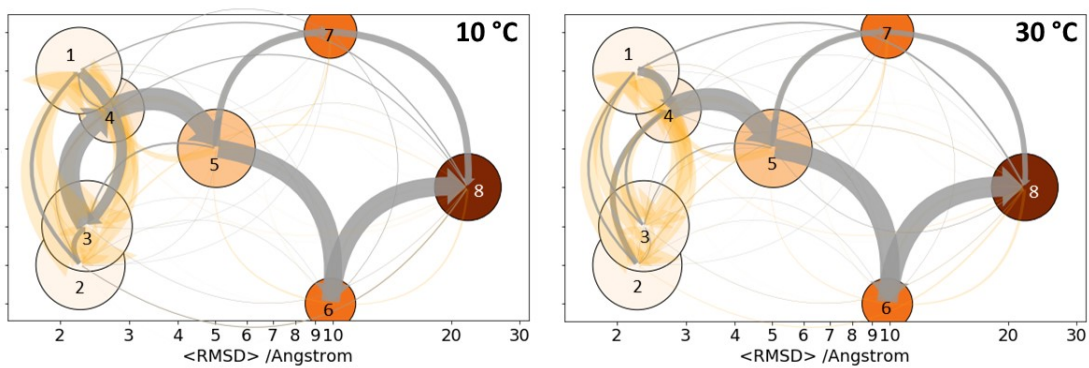

**Figure S6.** *Analysis of egress routes for benzene from T4L:L99A at 10 °C (left) and 30 °C (right) in RAMD trajectories. Clusters were defined by clustering of the last 300 frames in each trajectory in the IFP space; simulations at 10, 20 and 30 °C were included in the clustering procedure. Dissociation pathways are shown in a graph representation; each node represents a cluster that is colored and positioned according to increasing mean RMSD of benzene in the cluster from in the starting complex; the node size denotes the cluster population; transitions between nodes are indicated by arrows for simulations: the net transition flux between nodes is shown by gray arrows with their thickness proportional to the flux magnitude; transition between states are shown by orange arrows with their thickness proportional to the transition number.*

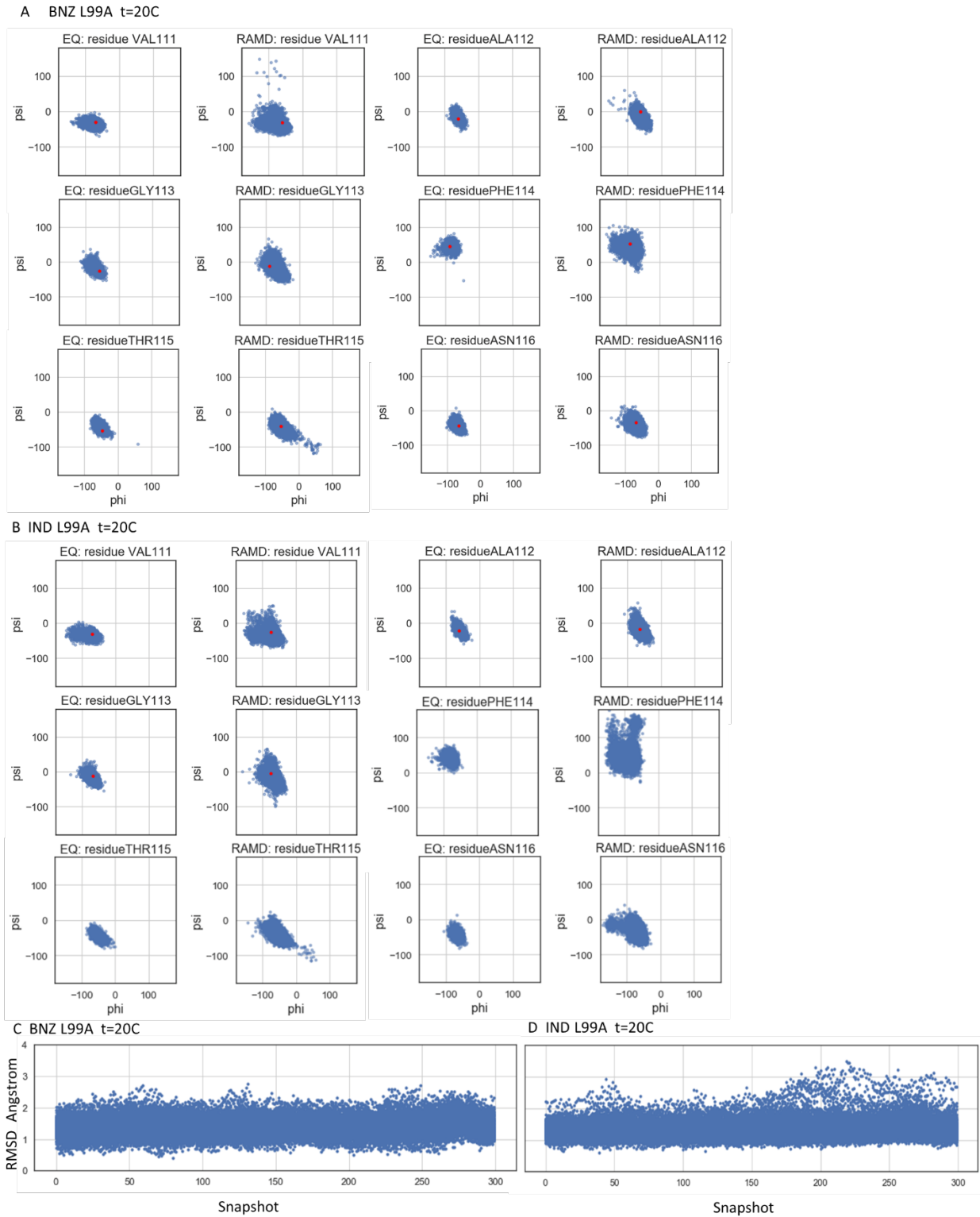

**Figure S7.** *Conformational changes in T4L:L99A upon ligand egress. (A, B) Ramachandran plots for several residues in the region between the F and G helices generated from the equilibration (EQ) and*

RAMD trajectories (the last 300 snapshots from each trajectory were used in the analysis) of T4L:L99A for the egress of (A) benzene and (B) indole at 20 °C. The red dots show the angles for the corresponding crystal structures. (C, D) RMSD of the non-hydrogen atoms of the residues of helix F (G108 – T115) relative to the starting structure of the T4L:L99A – ligand complex, plotted for the last 300 snapshots of each RAMD trajectory for (C) benzene and (D) indole dissociating from T4L:L99A at 20 °C. Conformational changes in T4L:L99A from a highly to a lowly populated state, characterized by NMR[41,42], were observed to facilitate benzene unbinding through path FGH in several computational studies[20,25], while in other studies, such changes were observed for the binding of bulkier ligands to T4L:L99A, but not for benzene[11]. This conformational change mainly involves helix F and one of its features is a change in the F114 psi angle from 50° to -40°. (A, C) show that there are no clear conformational changes of the helix F backbone or of F114 upon benzene dissociation that would characterize a transition to a lowly populated state of the protein with a conformation different from that in the bound structure although the protein RMSD and the dihedral angles for residues V111-T115 show greater variation in RAMD dissociation simulations than in equilibration MD. The variations in protein conformation are notably greater in the case of indole dissociation (B, D), which is in line with the results of Ref. [11] showing that protein distortion is caused by ligands larger than benzene.

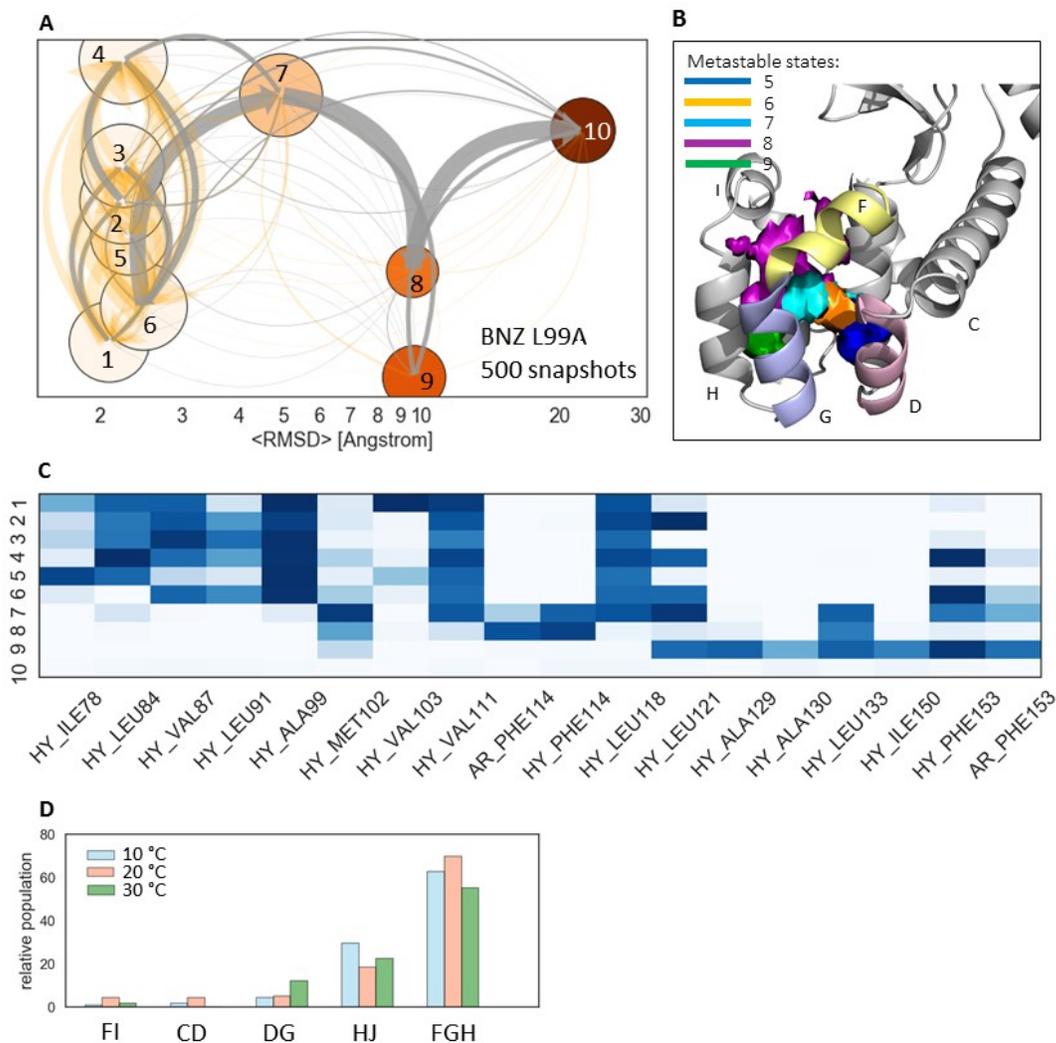

**Figure S8.** *Analysis of benzene unbinding from T4L:L99A in IFP space using the last 500 snapshots of each RAMD trajectory. Simulations at temperatures of 10, 20 and 30 °C and pH 5.5 were included in the clustering procedure. (A) Dissociation pathways are shown in a graph-representation; each node represents a cluster or metastable state that is colored and placed according to increasing mean RMSD of benzene in the cluster from in the starting complex; the node size denotes the cluster population; transitions between nodes are indicated by arrows for simulations at 20 °C: the net transition flux between nodes is shown by gray arrows with thickness proportional to the flux magnitude; transitions between states are shown by orange arrows with the thickness proportional to the transition number. (B) The clusters shown in (A) are displayed on the T4L:L99A structure; selected clusters are shown by isosurfaces of the ligand COM population density mapped onto a 3D grid; helices D, F and G are shown in pink, yellow and blue, and benzene is shown in ball and stick representation in cyan in its bound position. (C) Cluster composition,*

*shown as contacts between T4L:L99A and benzene and ordered by benzene RMSD. (D) Relative pathway populations observed in simulations at the three different temperatures of 10, 20 and 30 °C.*

# References

[27]  L. Liu, A. J. V. Marwitz, B. W. Matthews, S.-Y. Liu, *Angew. Chemie Int. Ed.* **2009**, *48*, 6817–6819.

[28]  E. Baldwin, W. A. Baase, X. Zhang, V. Feher, B. W. Matthews, *J. Mol. Biol.* **1998**, *277*, 467–485.

[29]  A. Morton, B. W. Matthews, *Biochemistry* **1995**, *34*, 8576–8588.

[30]  T. J. Dolinsky, J. E. Nielsen, J. A. McCammon, N. A. Baker, *Nucleic Acids Res.* **2004**, *32*, W665-7.

[31]  T. J. Dolinsky, P. Czodrowski, H. Li, J. E. Nielsen, J. H. Jensen, G. Klebe, N. A. Baker, *Nucleic Acids Res.* **2007**, *35*, W522–W525.

[32]  J. A. Maier, C. Martinez, K. Kasavajhala, L. Wickstrom, K. Hauser, C. Simmerling, K. E. Hauser, *J Chem Theory Comput* **2015**, *11*, 3696–3713.

[33]  J. Wang, R. M. Wolf, J. W. Caldwell, P. A. Kollman, D. A. Case, *J Comput Chem* **2004**, *25*, 1157–1174.

[34]  C. I. Bayly, P. Cieplak, W. D. Cornell, P. A. Kollman', *J. Phys. Chem* **1993**, *97*, 10269–10280.

[35]  W. D. Cornell, P. Cieplak, C. I. Bayly, P. A. Kollman, *J. Am. Chem. Soc.* **1993**, *115*, 9620–9631.

[36]  M. S. M. S. Gordon, M. W. M. W. Schmidt, *Chapter 41 – Advances in Electronic Structure Theory: GAMESS a Decade Later*, Elsevier, Amsterdam, **2005**.

[37]  X. W. and P. A. K. D.A. Case, V. Babin, J.T. Berryman, R.M. Betz, Q. Cai, D.S. Cerutti, T.E. Cheatham, III, T.A. Darden, R.E. Duke, H. Gohlke, A.W. Goetz, S. Gusarov, N. Homeyer, P. Janowski, J. Kaus, I. Kolossváry, A. Kovalenko, T.S. Lee, S. LeGrand, T. Luchko, R. Luo, B., **2014**.

[38]  J. C. Phillips, R. Braun, W. Wang, J. Gumbart, E. Tajkhorshid, E. Villa, C. Chipot, R. D. Skeel, L. Kalé, K. Schulten, *J. Comput. Chem.* **2005**, *26*, 1781–1802.

[39]  S. K. Lüdemann, V. Lounnas, R. C. Wade, *J. Mol. Biol.* **2000**, *303*, 797–811.

[40]  D. B. Kokh, B. Doser, S. Richter, F. Ormersbach, X. Cheng, R. C. Wade, *J. Chem. Phys.* **2020**, *153*, 125102.

[41]  G. Bouvignies, P. Vallurupalli, D. F. Hansen, B. E. Correia, O. Lange, A. Bah, R. M. Vernon, F. W. Dahlquist, D. Baker, L. E. Kay, *Nature* **2011**, *477*, 111–117.

[42]  F. A. A. Mulder, A. Mittermaier, B. Hon, F. W. Dahlquist, L. E. Kay, *Nat Struct Mol Biol* **2001**, *8*, 932–935.